\documentstyle[aps,preprint]{revtex}

\begin{document}
\title{Observation of Magnetic Flux Generated Spontaneously During a Rapid Quench
of Superconducting Films}
\author{A. Maniv, E. Polturak and G. Koren}
\address{Physics Department, Technion - Israel Institute of Technology, Haifa 32000,\\
Israel}
\date{\today}
\maketitle
\pacs{05.70.Ln, 74.40.+k, 11.27.+d}

\begin{abstract}
We report observations of spontaneous formation of magnetic flux lines
during a rapid quench of YBa$_{2}$Cu$_{3}$O$_{7-\delta }$ films through T$%
_{c}$. This effect is predicted according to the Kibble-Zurek mechanism of
creation of topological defects of the order parameter during a
symmetry-breaking phase transition. Our previous experiment, at a quench
rate of 20K/sec, gave null results. In the present experiment, the quench
rate was increased to 
\mbox{$>$}%
10$^{8}$ K/sec. Within experimental resolution, the dependence of the
measured flux on the cooling rate is consistent with the prediction.
\end{abstract}

\bigskip A certain class of grand unified theories describes the early
universe in terms of a series of symmetry breaking phase transitions. In
that context, Kibble$\ \cite{bib 1}$ predicted that if a system having a
complex order parameter is quenched through a phase transition into an
ordered state,topological defects will be created. This is due to the
evolution of uncorrelated regions of the newly formed phase, each region
having different values of the order parameter. The defects appear in
between several coalesced regions of this kind. Zurek$\cite{bib 2}$
developed this idea to predict the initial density of defects as a function
of quench rate and suggested specific experiments on condensed matter
systems to test this scenario. The natural candidates for such tests are
superfluids and superconductors, in which the topological defects are
quantized vortex lines. Superconductors have an added degree of complexity,
due to the presence of the gauge field A which evolves with time. Results of
various experiments done so far are not unambigous; spontaneously generated
vortices were observed in superfluid $^{3}$He$\cite{bib 8,bib 9}$ but not in 
$^{4}$He$\cite{bib 7}$. Experiments with homogeneous superconductors have so
far shown null results$\cite{bib 100}$. Other related systems are liquid
crystals undergoing an isotropic-nematic transition$\cite{bib 4,bib 5}$ (in
this case the topological defects are disclinations). Experiments done with
superconducting rings$\cite{bib 101}$ were done in a regime where usual
thermal fluctuations dominate, rather than Zurek's mechanism. Experiments
using Josephson junctions$\cite{bib 102,bib 103}$ gave results broadly
consistent with the Zurek scenario. However, these are intrinsically
inhomogeneous systems which do not fall into the class of systems directly
comparable with this theory. Here, we report the results of a new, improved
experiment with superconducting films$\cite{bib 100}$.

We used 300nm thick c-axis oriented YBa$_{2}$Cu$_{3}$O$_{7}$ films with T$%
_{c}\simeq $90K, grown on a SrTiO$_{3}$ substrate and patterned into a disc,
6mm in diameter .The basic experimental setup is described in Ref. 7.
Briefly, the sample is placed atop the sensing coil of a HTSC SQUID
magnetometer, at a distance of 1mm. In our arrangement the SQUID remains at
a temperature of 77K, and is not affected by the temperature of the sample
which can be heated and cooled independently. To avoid spurious magnetic
fields generated by electrical current used in resistive heating, the film
is heated above T$_{c}$ using a light source and cools by exchanging heat
with its environment. The system is carefully shielded from the earth's
magnetic field, with a residual magnetic field of less than 0.05 mG.
Additional small coil adjacent to the sample was used to test the field
dependence of the results. Instead of $\sim $ 1 sec long illumination from a
quartz lamp used to heat the sample in our previous work$\cite{bib 100}$,
the light source in the present experiment is a pulsed YAG\ laser$\cite{bib
107}$ . Single pulses, 10$^{-8}$ sec long, were used to heat the film. After
passing through a diffuser, the laser pulse passes through the substrate and
illuminates homogenously a 9mm diameter area of the film, larger than our
sample. At a laser wavelength of 1.06 $\mu $m, the SrTiO$_{3}$ substrate is
transparent and practically all the light is absorbed in the film. Hence,
only the film heats up, while the substrate remains near the base
temperature of 77K. The 1mm thick substrate has a heat capacity about 10$%
^{3} $ larger than that of the film. Therefore, an energy of $\sim $ mJ is
sufficient to heat the film above T$_{c}$, rather than a $\sim $ J used
previously$\cite{bib 100}$. The heat from the film escapes into the
substrate, which acts as a heat sink. This strongly reduced thermal mass
which is cooled allows us to achieve cooling rates in excess of 10$^{8}$
K/sec, 7 orders of magnitude faster than previously$\cite{bib 100}$. The
cooling rate at T$_{c\text{ }}$can be varied by changing the amount of
energy delivered by the laser pulse (see Fig. \ref{Fig 1} ). As the figure
shows, increasing this energy reduces the cooling rate. The cooling rate was
determined by monitoring the time dependence of the film resistance
following a laser pulse. Because heat flow into the substrate takes place in
a direction normal to the plane of the film, the temperature of the sample
is approximately the same along its lateral dimensions. Measurement of the
net flux is done continuosly during the heating-cooling cycle.

Achieving as high a cooling rate as possible is extremely important in order
to enter the regime where the Zurek scenario applies. To observe the effect,
the system needs to be out of equilibrium over a temperature interval wider
than the critical regime near T$_{c}$. Due to the anisotropy of the
superconducting properties of YBCO, our films are effectively 2D near T$_{c}$%
. We therefore expect that quantized vortices will develop only
perpendicular to the film surface. The 2D Ginzburg-Landau model yields a 10$%
^{-2}$K as the width of the critical regime$\cite{bib 105}$ (the 2D-XY model
gives an even smaller value). At quench rates between 10$^{7}$-10$^{8}$K/sec
the system remains out of equilibrium over an interval of 0.1-0.2K of T$_{c}$%
. Thus, the condition for observing this effect are satisfied in our case.
Because the coherence length of HTSC is small, the predicted $initial$\
flux-line density $n_{i}$ generated in the film by a thermal quench is very
large according to this scenario, $n_{i}=10^{11}-10^{12}cm^{-2}$\ . This
includes both vortices and antivortices, with the lower value corresponding
to G-L model, and the larger to 2D-XY model. In our experiment we measure
directly the difference between the number of vortices and antivortices,
namely the net flux. If the picture of regions having a well defined phase,
and with the choice of a minimal phase gradient between the regions (the
geodesic rule$\cite{bib 106}$) is correct, than the {\em net} rms flux
should scale as $n_{net}\sim (L)^{1/2}(n_{i})^{1/4}$, where L is the length
of the sample perimeter. In terms of the quench rate dT/dt, $n_{net}\sim
(dT/dt)^{1/8}$. We point out that this weak dependence leads to a predicted $%
n_{net}$ which inreases only by 20\% while the quench rate increases by an
order of magnitude. In the range of our experiment, the net flux density is
predicted to be $\sim $10$^{2}${\em \ }$\phi _{0}/$cm$^{2}${\em \ }.The
noise level of our magnetometer is equivalent to a flux noise of $\sim $5
net $\phi _{0}$, referred to the film. Thus, the effect should be
observable. It should be noted that our measuring system can detect only the
net flux ''frozen'' in the film, due to the fact that the film's total
cooling time is of the order of 1$\mu \sec $, while the SQUID system
responds on a time scale of about 10 $\mu \sec $. Flux will be ''frozen'' in
the film if the pinning site density is much larger than the flux density.
The pinning site density in similar films was estimated in Ref. 17 (and in
references therein) as 1-6x10$^{10}$\ cm$^{-2}\gg $ $n_{net}$. Since pinning
inYBCO films is very strong at temperatures below the critical regime, we
conclude that the net flux generated during the quench should remain inside
the film.

{\it \bigskip }

In a typical experiment, the SQUID's output is recorded vs. time as the
sample cools following a thermal quench. Such measurements were performed
both on superconducting samples and on a control sample. The control sample
is a similar film of underdoped YBCO, having T$_{c}$ of 60K, which is not
superconducting in our temperature range. Indeed, net flux was observed with
the superconducting film while no flux was seen in the control experiment.
Fig. \ref{Fig 2} shows data from 100 such measurements. According to the
Zurek scenario, the rms of spontaneously generated flux should increase in
amplitude with the cooling rate while the sign of the net flux should be
random from one quench to the next. Fig. \ref{Fig 2} clearly shows that this
indeed is the case. Further, the signal\ obtained with superconducting film
is much larger than the control signal. A typical distribution of flux from
such measurements is plotted in Fig. \ref{Fig 3}. It can be seen that the
distribution of the signal is symmetric about zero flux, as expected from
this scenario. In order to check for the effect of any residual field, our
measurements were repeated under different fields up to 10mG, about 10$^{3}$
times larger than our residual field. The inset in Fig. \ref{Fig 3} shows
that the magnetic field has no significant influence on our results.

\bigskip

The net flux was deduced by deconvoluting the noise from the measured
signal. The net flux distribution width vs. the cooling rate is shown in
Fig. \ref{Fig 4}. The solid line shows the $(dT/dt)^{1/8}$ dependence
predicted by Zurek$\cite{bib 2,bib 3}$, with a prefactor given by Rudaz et
al.$\cite{bib 106}$. To agree quantitatively with the data, the theoretical
prediction was scaled down by a factor of 4. It is seen that the data are
consistent with this prediction. The consistency of our results with the $%
(dT/dt)^{1/8}$ dependence further implies that the geodesic rule is valid in
a non equilibrium regime. The validity of this rule was not considered
obvious$\cite{bib 12}$. Extrapolating the data shown in Fig. \ref{Fig 4}
down to a cooling rate of 20K/sec, that of ref. \cite{bib 100}, gives a
predicted flux density of $\sim $ 6 $\phi _{0}$/cm$^{2}$ for that
experiment. This value is very close to the noise level, and thus explains
the null result obtained\cite{bib 100}.  Finally, we carried out several
checks to see how the magnitude of a signal is influenced by temperature
gradients. The presence of temperature gradients is important with respect
to the homogeneous approximation$\cite{bib 3,bib 24}$. We estimate our
maximum temperature gradient as $\nabla T\sim $ 1 K/cm parallel to the film
surface, similar to the spread of \ T$_{c}$ across the film. Under these
conditions, the homogeneous approximation is valid in our experiment. In
these additional experiments, we created intentional temperature gradients
in order to check whether the scaling factor between theory and experiment
cited here is a result of such gradients present in our film. We found that
this was not the case$\cite{bib 108}$.

We thank M. Ayalon, L. Yumin, and S. Hoida for technical assistance. We
thank P. Leiderer, B. Biehler and B. Shapiro for useful discussions. This
work was supported in part by the Israel Science Foundation, The Hertz
Minerva Center on Superconductivity, and the Technion Fund for Research.

\bigskip

\begin{figure}[tbp]
\caption{Temperature cycle of the film after a laser pulse. The energy of
the pulse is 7.1 mJ (closed symbols) and 3.1 mJ (open symbols). The
horizontal line is T$_{c}$ for our sample. Note that the cooling rate
through T$_{c}$ is slower for the high energy pulse.}
\label{Fig 1}
\end{figure}

\begin{figure}[tbp]
\caption{ Typical sequence of 100 consecutive magnetometer readings each
following a separate quench. Open symbols- control sample, closed
symbols-superconducting film. The lines connect successive data points.}
\label{Fig 2}
\end{figure}

\begin{figure}[tbp]
\caption{ Typical histogram of spontaneous flux from several hundred
quenches. The solid line is Gaussian fit. The inset shows that the
distribution width does not depend on the external field.}
\label{Fig 3}
\end{figure}

\begin{figure}[tbp]
\caption{Dependence of the distribution of spotaneous flux vs. the cooling
rate. The solid line is the prediction of ref. [2],[15] scaled to fit the
data. .}
\label{Fig 4}
\end{figure}

\bigskip

\bigskip

\bigskip

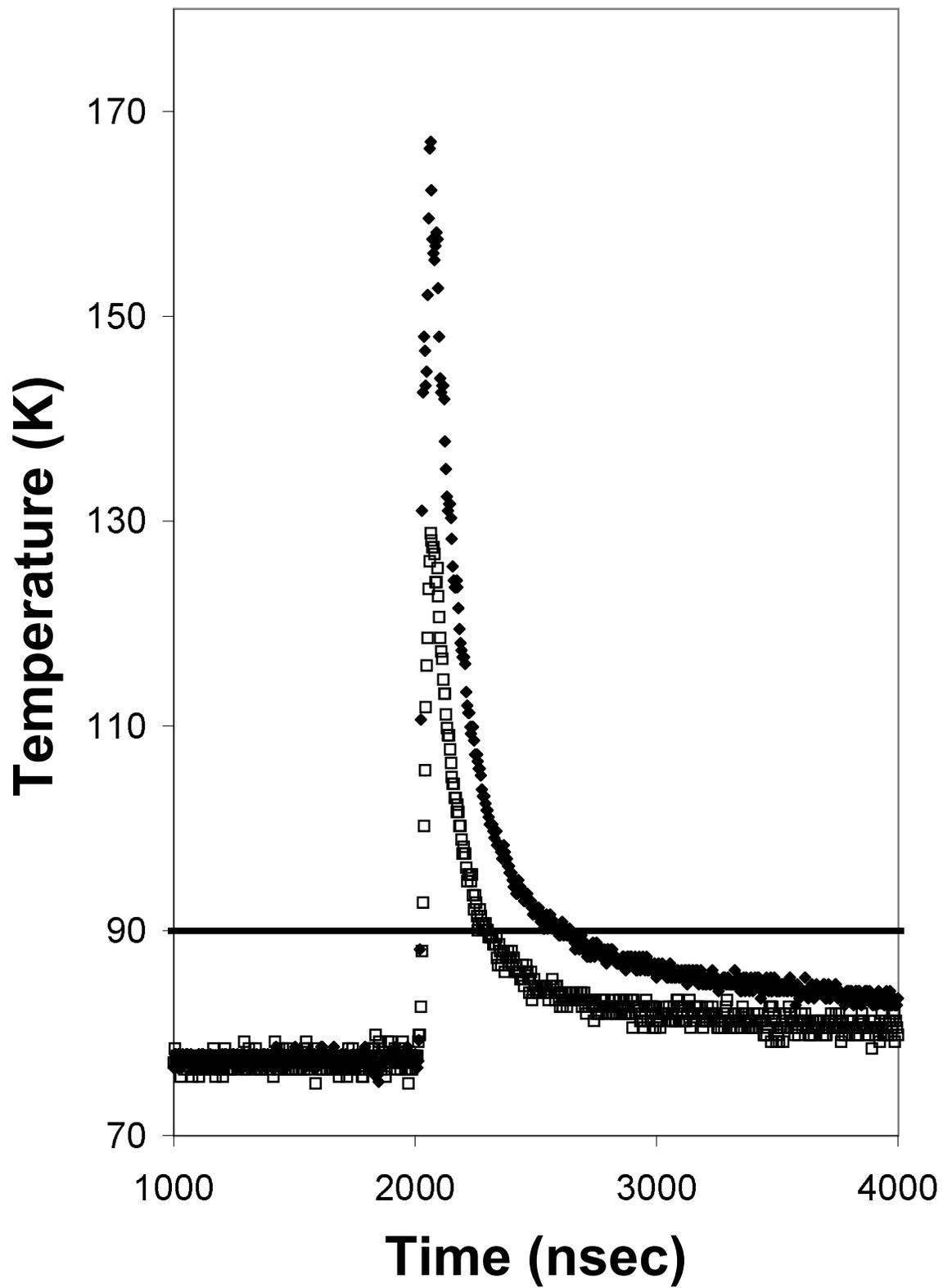
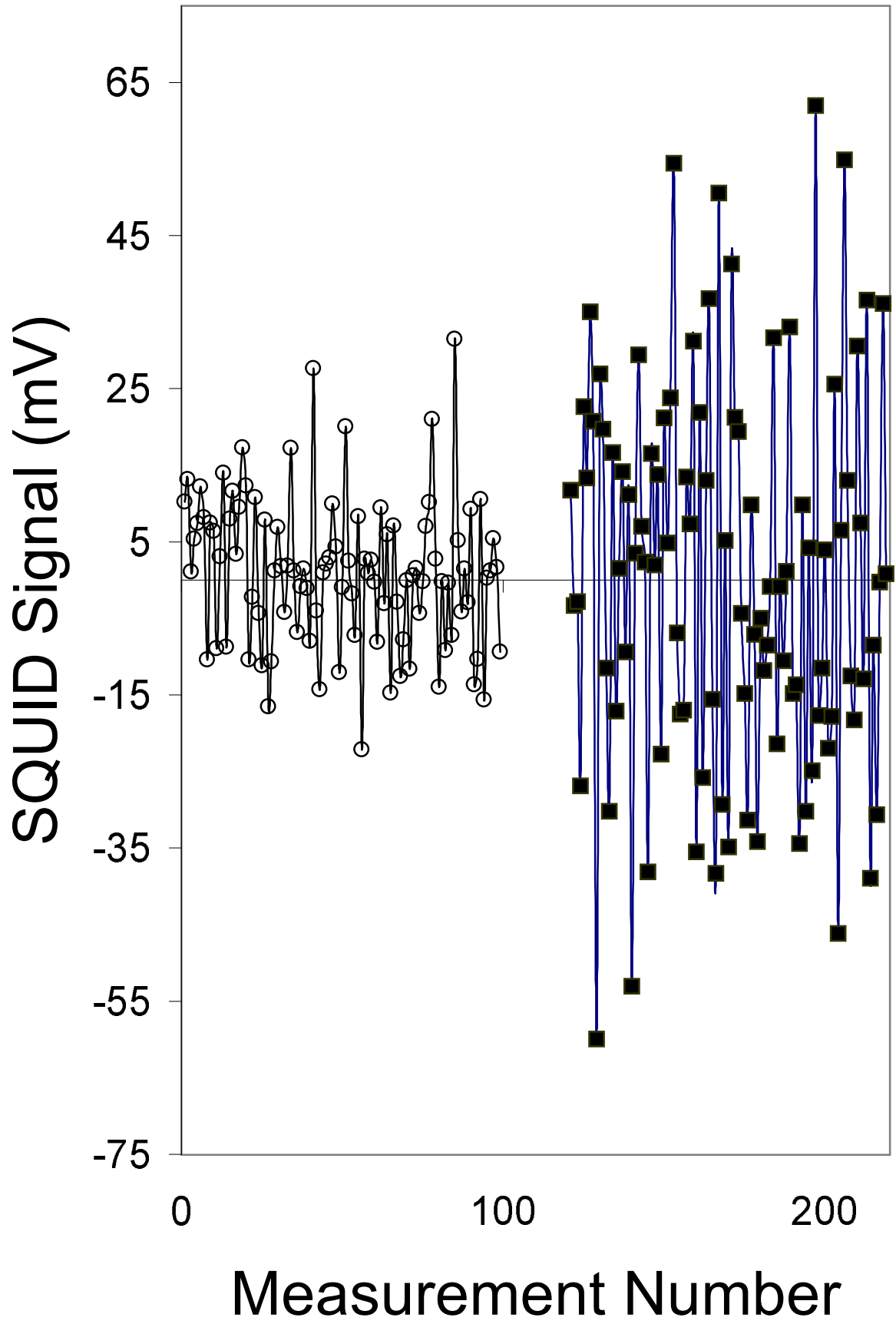
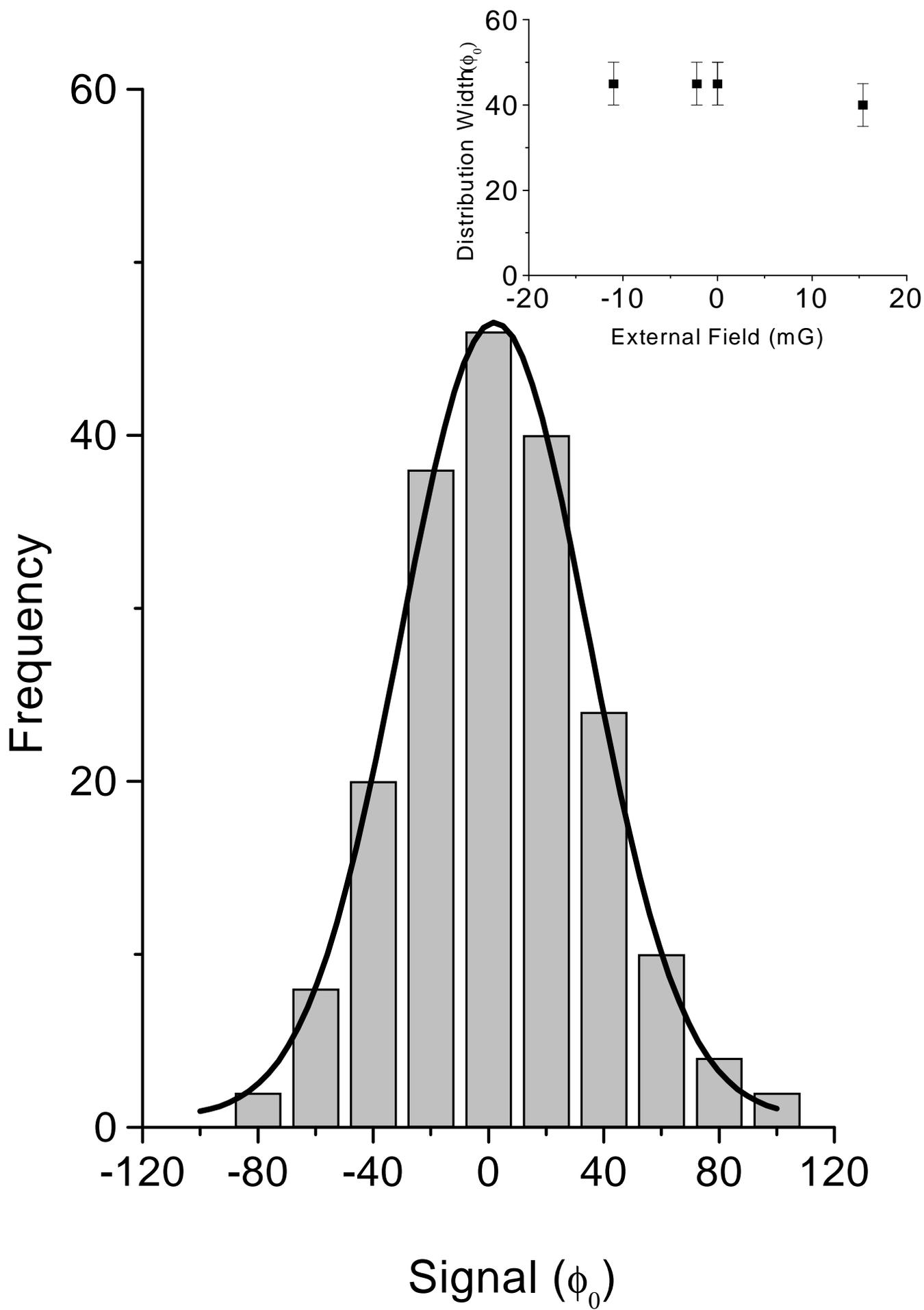
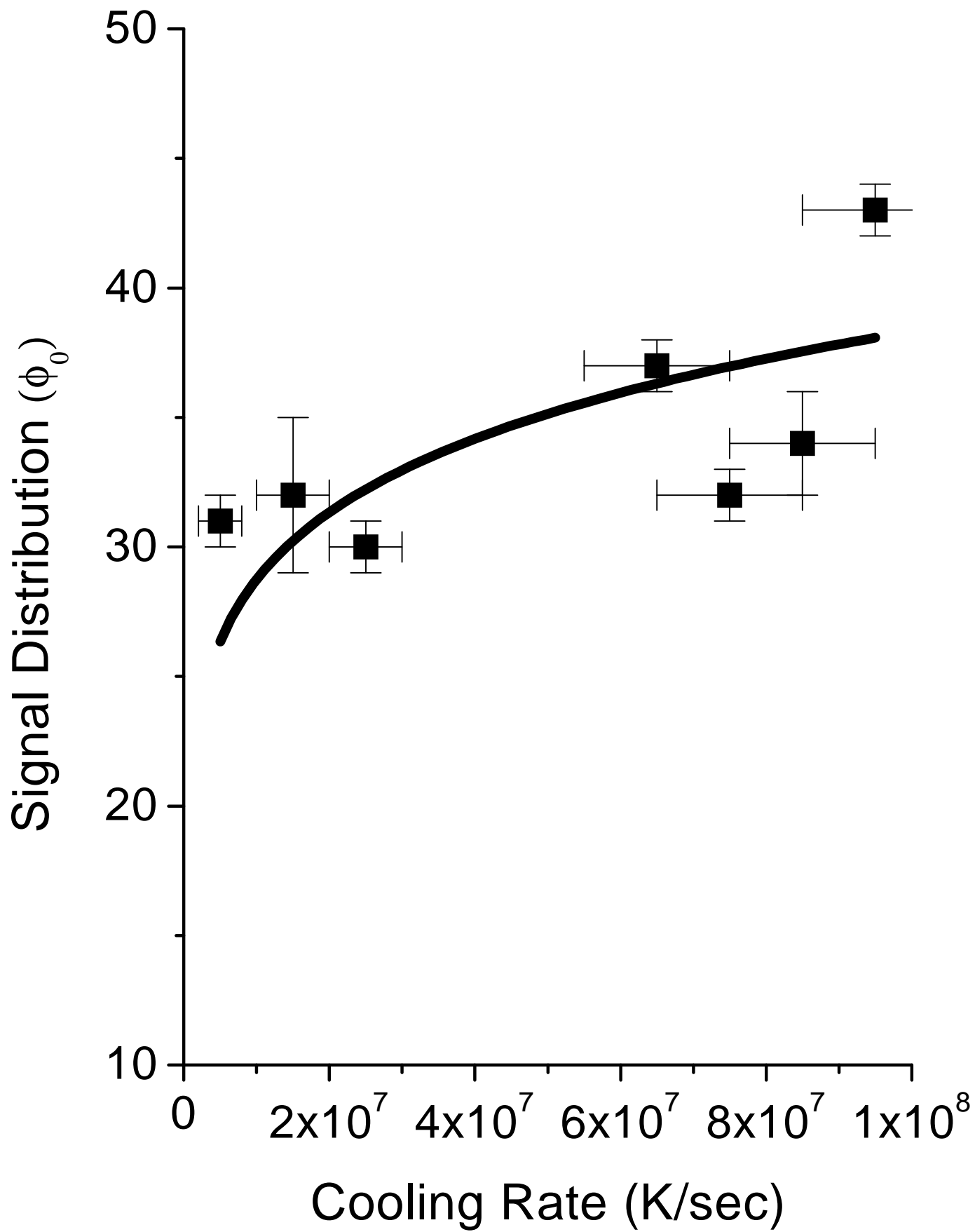
\end{document}